\documentstyle[12pt]{article}

\begin{document}
\vspace*{1.0cm}
\begin{flushright} {\bf ISU-IAP.Th96-05,\ Irkutsk\\
                         } \end{flushright}
\vspace*{1.5cm}
\begin{center}
{\Large \bf Effect of the unitary mixing scalar--vector.}\\[2.5cm]
{\Large  A.E.Kaloshin}
\end{center}
\begin{center}
{\it Institute of Applied Physics, Irkutsk State University,
664003 Irkutsk, Russia.
\footnote{\small E-mail: kaloshin@physdep.irkutsk.su}\\[1cm] }
\end{center}
\vspace{1cm}
{\large Abstract}
\vspace{0.5cm}

We consider the procedure of dressing of scalar and vector
particles when there exists the off--diagonal loop
connecting  vector and scalar propagators.
Instead of single Dyson equations for scalar and vector,
we have in this case a system
of three equations for coupled full propagators.
Using the $\pi-a_1$ system as an example,
we discuss the physical meaning of this effect and the
renormalization procedure for coupled propagators.
The considered effect of the unitary S~--~V
mixing may exits also in the "Higgs boson~--~Z~(~W~)"
system for extended electroweak models.

\newpage
{\large 1.\ Introduction.\\[0.5cm]}

The problem of the vector boson "dressing" (~i.e. its
turning into the finite width particle~) was discussed
last years in different aspects,
see e.g.~\cite{Castro1,Nowak,Veltman,Atwood}.
It was initiated first of all by experiments on the W and Z
production, where the more precise  measurements allow to
investigate rather subtle effects including radiative corrections.
The main theoretical problems are related here with
a gauge invariance, that is, how to introduce the mass and width
of unstable gauge boson in a gauge-invariant manner.

At present work we discuss rather specific effect
which can take place at dressing of a vector boson.
It turns out that in some cases, when the vector current is not
conserved, there appears a loop  connecting
vector and scalar propagators.
It does not contradict with the angular momentum
conservation because a closer look shows that
such loop relates scalar propagator only with longitudinal part of
vector propagator. As it is well--known, the longitudinal part
corresponds to additional scalar degree of freedom which
is contained in  vector field. This degree of freedom
is eliminated usually under quantization by some extra condition,
for example, $\partial_{\mu}V^{\mu}$~=~0.
But with accounting of loops (~i.e. quantum effects~) this degree
of freedom turns again into the "game" and it leads to a quite unusual
form of full propagators. As a result, the joint efforts of
scalar and the longitudinal part of vector propagator
create an effective scalar resonance.
In particular, such effect takes place in hadron physics in
the system $\pi~-~a_1 $ (~or $\pi^{\prime} - a_1 $~).

The considered effect of unitary  S~--~V mixing can arise also
for massive gauge bosons and  in this case
the Higgs scalars play the role of
partners for W and Z in a joint dressing.
But in the Standard Model, where  only one scalar neutral
Higgs particle exists, this phenomenon is absent,
the corresponding off--diagonal loop is equal to zero.
Nevertheless, this effect can arise in extended electroweak
models, where charged or (and) pseudoscalar
Higgs particles exist.\\
\newpage

{\large 2.\ Formalism of S~--~V mixing.\\[0.2cm]}

Let us consider bare vector propagator in unitary
gauge:
\footnote{The unitary gauge is  conventionally used in discussion
of dressing and we also use it for simplicity. The S~--~V
mixing may be considered in general $\xi$ -- gauge but
there appears an additional scalar taking part in game -- so called
Higgs ghost. }
\begin{equation}
P^{\mu\nu} = i\  \frac{- g^{\mu \nu} + p^{\mu} p^{\nu}/ M^2}
{p^2 - M^2 + i\ \epsilon} = i\ \{ \rho^{\mu\nu}_T\cdot \pi_T(p^2) +
\rho^{\mu\nu}_L\cdot \pi_L(p^2)\} ,
\end{equation}
where
\begin{eqnarray}
\rho^{\mu\nu}_T = - g^{\mu \nu} + p^{\mu} p^{\nu}/ p^2,\ \ \ \
\rho^{\mu\nu}_L = p^{\mu} p^{\nu}/ p^2 \nonumber\\
\pi^T(p^2) = \frac{1}{p^2 - M^2 + i\ \epsilon},\ \ \ \
\pi^L(p^2) = \frac{1}{M^2} .
\end{eqnarray}

To obtain a full propagator from a bare one, we should
write the Dyson equation, which summarizes the self--energy
inserts, and this equation determines the full propagator.
But since vector propagator has tensor structure,
transverse and longitudinal parts will dress in
different ways. It is convenient to project the vector loop onto
T and L components.
\begin{equation}
J^{\mu\nu} = \rho^{\mu\nu}_T\cdot J^T(p^2) +
\rho^{\mu\nu}_L\cdot J^L(p^2)
\end{equation}
Then we have the standard answer for full unrenormalized
propagators:
\begin{eqnarray}
\pi^T(p^2) = \frac{1}{p^2 - M^2 + i\ \epsilon} \ \
&\Rightarrow& \Pi^T(p^2) = \frac{1}{p^2 - M^2 + J^T(p^2)}
\nonumber\\
\pi^L(p^2) = \frac{1}{M^2} \ \
&\Rightarrow& \Pi^L(p^2) = \frac{1}{M^2 + J^L(p^2)}
\label{SA}
\end{eqnarray}

Another situation will arise if there exists the suitable by
quantum numbers scalar particle, which dresses by the same
intermediate state and, besides, there occurs the off--diagonal loop
connecting the scalar and vector propagators.
It is evidently  from the beginning that
such transition loop may relate a scalar propagator
only with the longitudinal part of vector one.
If the off--diagonal loop really exists then besides the full
scalar and vector propagators a new object arises:
a full off--diagonal propagator
which relates the scalar and vector vertices.

As a result instead of one Dyson equation we have
the system of three equations of the following form
\begin{eqnarray}
\begin{array}{ccc}
\Pi_{11} &=& \pi_{11} -
\Pi_{11}\  J_{11}\ \pi_{11} -
\Pi^{\alpha}_{12}\ i J^{\alpha}_{21}\  \pi_{11}
\\
\Pi^{\mu}_{12} &=&
 \ \ \ \ - \Pi_{11}\ i J^{\alpha}_{12}\ \pi^{\alpha\mu}_{22} -
\Pi^{\alpha}_{12}\ J^{\alpha\beta}_{22}\
\pi^{\beta\mu}_{22}
\\
\Pi^{\mu\nu}_{22} &=& \pi^{\mu\nu}_{22} -
\Pi^{\mu}_{21}\ i J^{\alpha}_{12}\ \pi^{\alpha\nu}_{22} -
\Pi^{\mu\alpha}_{22}\ J^{\alpha\beta}_{22}\
\pi^{\beta\mu}_{22}.
\end{array}
\label{WS1}
\end{eqnarray}

Here we introduced  a matrix notations for scalar (11),
vector (22), transition (12) propagators and similar
for loops, the factor~i in~$J^{\alpha}_{12}$ is written
for convenience. If $J^{\alpha}_{12}~=~0$, we have the standard
situation of single equations.
For off--diagonal propagators and loops
we can pass over to corresponding scalar functions.
\begin{eqnarray}
\Pi^{\alpha}_{12}(p) = p^{\alpha}\ \Pi_{12}(p^2),\ \ \ \
J^{\alpha}_{12}(p) = p^{\alpha}\ J_{12}(p^2),\nonumber \\
\Pi^{\alpha}_{21}(p) = - \Pi^{\alpha}_{12}(p),\ \ \ \
J^{\alpha}_{21}(p) = - J^{\alpha}_{12}(p)
\end{eqnarray}

The last equation in the system~(\ref{WS1})
should be projected onto transverse and longitudinal
components. As a result we obtain the single equation
for transverse component
\begin{equation} \Pi_{22}^T = \pi_{22}^T - \Pi_{22}^T \
J_{22}^T\ \pi_{22}^T
\end{equation}
and the transverse part remains as before (\ref{SA}).
As for the longitudinal component, it enters
the system of three equations.
\begin{eqnarray} \begin{array}{ccc} \Pi_{11}
&=&  \pi_{11} - \Pi_{11}\ J_{11}\ \pi_{11} + p^2\ \Pi_{12}\ i J_{21}\
\pi_{11} \\
\Pi_{12} &=& - \Pi_{11}\ i J_{12}\ \pi^L_{22} - \Pi_{12}\
J^L_{22}\ \pi^L_{22} \\
\Pi^L_{22} &=& \pi^L_{22} + p^2\ \Pi_{21}\
i J_{12}\ \pi^L_{22} - \Pi^L_{22}\ J^L_{22}\ \pi^L_{22} \end{array}
\label{WS2}
\end{eqnarray}

Solution of this system in most general form is
\begin{eqnarray}
\Pi_{11}(p^2) = \frac{(\pi^L_{22})^{-1} + J^L_{22}}{D(p^2)},\ \
\Pi_{12}(p^2) = - \frac{i J_{12}(p^2)}{D(p^2)},\ \
\Pi_{22}^L(p^2) = \frac{(\pi_{11})^{-1} + J_{11}}{D(p^2)},
\nonumber\\
D(p^2) = \left[ (\pi_{11})^{-1} + J_{11} \right]\
\left[ (\pi^L_{22})^{-1} + J_{22} \right]\ -
p^2 \ \left[ J_{12}(p^2) \right] ^2 .
\label{Sol}
\end{eqnarray}

Note that similar formulae ("resonances with unitary mixing")
are well--known in  hadron physics, see, e.g.~\cite{ADS},
in application to scalar resonances. They provide the unitary
condition in the case of few resonances with the same quantum
numbers.

So with accounting of S--V mixing the dressed
propagators take the form:
\begin{equation}
\Pi^T_{22}(p^2) = \frac{1}{p^2 - M^2 + J^T_{22}(p^2)}
\end{equation}
\begin{equation}
\Pi_{11}(p^2) = \frac{M^2 + J^L_{22}(p^2)}{D(p^2)},\ \
\Pi_{12}(p^2) = - \frac{i J_{12}(p^2)}{D(p^2)},\ \
\Pi_{22}^L(p^2) = \frac{p^2 - \mu^2 + J_{11}(p^2)}{D(p^2)}.
\label{Answer}
\end{equation}
\begin{equation}
D(p^2) = \left[ p^2 -\mu^2 + J_{11}(p^2) \right]\
\left[ M^2 + J_{}^L(p^2) \right]\ -
p^2 \ \left[ J_{12}(p^2) \right] ^2\ \ \ \ \ \ \ \ \ \ \ \
\label{D}
\end{equation}
Here M and  $\mu$  are bare masses of vector and scalar
particles.

The obtained formulae have rather clear meaning. The loop
connecting scalar and vector propagator is in fact the
transition of scalar
particle again to scalar "particle" contained in
vector field~$V^{\mu}$. For free vector field this degree
of freedom is usually removed by some extra condition
providing the self--consistent quantization. But with accounting
of loop effects this degree of freedom appears again
in the form of mixing of the longitudinal part of vector
propagator with scalar one.
The necessary condition for such effect
is the non--conservation of the corresponding vector current
$\partial_{\mu}J^{\mu} \not= 0$.\\[0.3cm]

{\large 3.\ \ The $\pi - a_1$ system.\\[0.3cm]}

The described situation of the unitary S -- V mixing
takes place in particular in the system of $\pi$ and $a_1$
mesons which have couplings with the three--pion
system. It is generally believed that 3$\pi$ system is saturated
by the quasi--two--particle states $\pi \sigma$ and $\pi \rho$.
We shall consider below the simplest case of a joint
$\pi$ and  $a_1$ dressing by  $\pi \sigma$ state.\\

Let us define the vertices in momentum space.\\[0.3cm]
\underline{ $ a_1(P) \to \pi(k)\ \sigma(q)$} \ \ \ vertex
with vector index $\mu$ in diagram has the factor\\
\hspace*{4cm} $(-1)\ g_{a_1\pi\sigma}\ (k -q)^{\mu}$,\\[0.2cm]
\underline{$\pi(P) \to \pi(k)\ \sigma(q)$} \ \ \ vertex is
\  \ \ $i\ g_{\sigma\pi\pi}$. \\[0.2cm]

We shall present the results of calculations below, where
$\mu$ and M are bare masses of $\pi$ and $a_1$ mesons,
and m is the $\sigma$--meson mass.\\
\underline{The transition loop $\pi - a_1$}
differs from zero
\footnote{Note that more general form of $a_1 \pi \sigma$ vertex
(~adding of total derivation term~) can not lead to zero answer
for transition loop. A nonzero contribution arises also from
the $\pi \rho$ intermediate state.}
and is of the form:
\begin{equation}
J_{12}^{\mu}(p) = i\ g_{a_1\pi\sigma} \ g_{\sigma\pi\pi}
\int \frac{d^4 l}{(2\pi)^4}\
\frac{(2l-p)^{\mu}}{(l^2 - \mu^2)((l-p)^2 - m^2)}
=  p^{\mu}\ J_{12}(p^2) ,
\label{loop12}
\end{equation}
where
\begin{equation}
J_{12}(p^2) = \frac{g_{a_1\pi\sigma} \ g_{\sigma\pi\pi}}{16\pi}
(m^2 -\mu^2)\
\frac{1}{\pi}\ \int \frac{ds}{s(s - p^2)}\cdot
\left( \frac{\lambda(s, m^2, \mu^2)}{s^2} \right)^{1/2} .
\end{equation}
Here $\lambda$ is the well--known function
$\lambda(a,b,c) = (a - b  - c)^2 - 4 b c$,\
limits of integration here and below are not shown,
they are evident.\\
\underline{Loop in the $a_1$ - propagator.}\\
\begin{equation}
J_{22}^{\mu\nu}(p) = -i\ g_{a_1\pi\sigma}^2
\int \frac{d^4 l}{(2\pi)^4}\
\frac{(2l-p)^{\mu} (2l-p)^{\nu}} {(l^2 - \mu^2)((l-p)^2 - m^2)} =
g^{\mu\nu}\cdot A(p^2) + p^{\mu} p^{\nu}\cdot B(p^2)
\end{equation}
Discontinuities of  A and B functions are:
\begin{eqnarray}
\Delta A(p^2) &=& -i\ \frac{g_{a_1\pi\sigma}^2}{24\pi}\
p^2\ \left( \frac{\lambda}{p^4}\right)^{3/2},
\nonumber\\
\Delta B(p^2) &=& i\ \frac{g_{a_1\pi\sigma}^2}{24\pi}\
\left[ \frac{\lambda}{p^4} + \frac{3 (m^2 - \mu^2)^2}{p^4} \right]
\left( \frac{\lambda}{p^4} \right)^{1/2},\nonumber\\
\lambda &=& \lambda(p^2,m^2,\mu^2) .
\end{eqnarray}

To restore the analytical function through its discontinuity, we need
one subtraction for A and two subtractions for B. After that the
transverse and longitudinal components are easily calculated.
\footnote{In such a way the poles $1/p^2$ in propagator
are canceled automatically from the all following expressions.}
\begin{equation}
J^T_{22} = -A(p^2),\ \ \ \ \ \ J^L_{22} = A(p^2) + p^2 B(p^2)
\end{equation}
That gives:
\begin{eqnarray}
J^T_{22}(p^2) &=& -A(0) - p^2 A^{\prime}(0) +
\frac{g_{a_1\pi\sigma}^2}{48\pi} \
\frac{p^4}{\pi}\ \int \frac{ds}{s (s - p^2)}
\left( \frac{\lambda}{s^2} \right)^{3/2}\\
J^L_{22}(p^2)&=&A(0) + p^2 [A^{\prime}(0) + B(0)] +
\frac{g_{a_1\pi\sigma}^2}{16\pi} (m^2 - \mu^2)^2\
\frac{p^4}{\pi} \int \frac{ds}{s (s - p^2)}
\frac{1}{s^2} \
\left( \frac{\lambda}{s^2} \right)^{1/2}\nonumber .
\label{loop22}
\end{eqnarray}
\underline{Loop in the $\pi$-- propagator.}\\
\begin{eqnarray}
J_{11}(p^2) &=& -i\ g_{\sigma\pi\pi}^2
\int \frac{d^4 l}{(2\pi)^4}\
\frac{1}{(l^2 - \mu^2)((l-p)^2 - m^2)}
\nonumber\\
&=& J_{11}(0) +
\frac{g_{\sigma\pi\pi}^2}{16\pi}
\cdot \frac{p^2}{\pi}\ \int \frac{ds}{s (s - p^2)}
\left( \frac{\lambda(s,m^2,\mu^2)}{s^2} \right)^{1/2}
\label{loop11}
\end{eqnarray}

One can see that the loop integrals $J_{11}$, $J_{12}$ É $J_{22}^L$,
involved into the system of equations  may be expressed by the
same function
\begin{equation}
H(p^2) =
\frac{1}{\pi}\ \int \frac{ds}{s (s - p^2)}
\left( \frac{\lambda(s,m^2,\mu^2)}{s^2} \right)^{1/2}
\end{equation}
These loops have the form (~with minimal subtractions~)
\begin{eqnarray}
J_{11}(p^2) &=& J_{11}(0) + g_1^2\cdot p^2 H(p^2)
\nonumber\\
J_{12}(p^2) &=& g_1 g_2 \cdot H(p^2)
\nonumber\\
J_{22}^L(p^2) &=& E + p^2 F + g_2^2 \cdot H(p^2),
\label{loops}
\end{eqnarray}
where we introduced the notations
$g_1^2 = g_{\sigma\pi\pi}^2/16\pi$ ,
$ g_2^2 = (m^2 - \mu^2)^2\cdot g_{a_1\pi\sigma}^2/16\pi $ and
\begin{equation}
E = A(0) - g_2^2 H(0),\ \ \ \ \ \
F = A^{\prime}(0) + B(0) - g_2^2 H^{\prime}(0)
\label{const}
\end{equation}
Here $H(0)$ and $H^{\prime}(0)$ are known quantities.

Let us consider the function $D(p^2)$ in
denominator of full propagators.
\begin{eqnarray}
D(p^2)&=&\left[ p^2 -\mu^2 + J_{11}(p^2) \right]\
\left[ M^2 + J_{22}^L(p^2) \right]\ -
p^2 \ \left[ J_{12}(p^2) \right]^2 = \nonumber\\
&=&[p^2 - \mu^2 + J_{11}(0)]\ [M^2 + E + p^2 F] + \nonumber\\
&&+ H(p^2)\ \{ p^2\ g_1^2\ (M^2 + E + p^2 F) +
g_2^2\ [p^2 - \mu^2 + J_{11}(0)]\ \}
\end{eqnarray}
Note that products of loops are canceled.
The similar cancelation arises for mixing of resonances
with the same spin~\cite{ADS}. But if to consider a few
intermediate states then such cancelation of all terms of the
$g^4$ order takes place only at some relations between coupling
constants.\\[0.2cm]
\underline{\large Renormalization.}\\

We shall renormalize propagators by means of the subtraction
on the mass shell,  supposing that in above formulae
$\mu$ and M  are the renormalized masses of pion and $a_1$.

First of all let us renormalize the transverse part of vector
propagator which has the Breit--Wigner form.
Normalization upon the total width requires to subtract $J_{22}^T$
twice on the mass shell.
\begin{equation}
Re\ J_{22}^T (M^2) = Re\ J_{22}^{T\ \prime} (M^2) = 0
\label{ms}
\end{equation}
That defines the constants A(0) and $A^{\prime}(0)$ .

In the case of few coupled propagators the procedure of subtraction
on a mass shell becomes somewhat different.
A zero of the function D($p^2$)  gives rise to pole of the
full pion propagator. If we require the unit residue
of the renormalized pion propagator then
from the first equation of the system (\ref{WS2})
we have the manifest condition :
\begin{equation}
\Delta(p^2) \equiv \Pi_{11}(p^2)\ J_{11}\ (p^2)  - p^2 \Pi_{12}(p^2)
\ i J_{12}(p^2) = 0
\ \ \ \ \ \ \mbox{at} \ \ p^2 = \mu^2 .
\label{26}
\end{equation}
It means that  sum of all inserts into the
external pion line is equal to zero. Using the solutions
(\ref{Answer}), one can obtain for the function $\Delta(p^2)$:
\begin{equation}
\Delta(p^2) = \frac{X(p^2)}{D(p^2)} =
\frac{X(p^2)}{(p^2 -\mu^2)(M^2 + J_{22}^L(p^2)) + X(p^2)},
\end{equation}
where
\begin{equation}
X(p^2) = J_{11}(p^2)\ (M^2 + J_{22}^L(p^2)) -
p^2 \ \left[ J_{12}(p^2) \right]^2 .
\label{X1}
\end{equation}
One can see from this expression that the requirement~(\ref{26})
is equivalent to the following conditions:
\begin{equation}
X(\mu^2)\ = \ X^{\prime}(\mu^2) = 0 .
\label{X2}
\end{equation}

In principal these requirements may be realized in different
ways. But there exists the natural physical condition on the
longitudinal part of vector propagator: it should not
have the pion pole after renormalization. So in the following
we shall accept this anzats for renormalization procedure:
{\sl
\begin{eqnarray}
\mbox{ The  longitudinal part of vector propagator}
\nonumber\\
\mbox { should not get a pole of scalar particle.
}
\label{anzats}
\end{eqnarray}
}
It leads to the following requirements for loops
\footnote{One can check by direct calculation that it is the only
possibility to satisfy~(\ref{X2}) at least
at small coupling constant $g_{a_1\pi\sigma}$ .}
(see~(\ref{Answer}),(\ref{X2})):

\begin{equation}
 J_{11}(\mu^2) = J_{11}^{\prime}(\mu^2) = 0,\ \ \ \ \ \ \
J_{12}(\mu^2) = 0
\label{Cond}
\end{equation}
As a result  $D(p^2)$ in vicinity of $p^2~=~\mu^2$
is of the form:
\begin{equation} D(p^2) = (p^2
-\mu^2)\ (M^2 + J_{22}^L(p^2)) + O((p^2 - \mu^2)^2)
\end{equation}
One can see from (\ref{Answer}), (\ref{Cond}) that only
the full pion propagator $\Pi_{11}$ has a pole
$1/(p^2~-~\mu^2)$, as for $\Pi_{12}$ and $\Pi_{22}^L$ --
they do not have it.

Note that after the mass renormalization for $a_1$ and $\pi$
the longitudinal loop is not defined completely.
\begin{equation}
J_{22}^L(p^2) = E + p^2\ F + g_2^2\ H(p^2)
\label{long}
\end{equation}
Recall that here the parameter F is still arbitrary.
But if in (\ref{long}) $F \not= 0$ then
$D(p^2)$ necessarily has a zero in complex plane at one
of the Riemann sheets, and we have put this parameter to zero
not to get poles in propagators  besides  the $p^2 = \mu^2$ one.

So we defined completely the full renormalized propagators
with account of the loop S--V transitions. The final formulae
are collected in Appendix.\\[0.3cm]

{\large 4.\ \ The $\pi^{\prime} - a_1$ system.\\[0.3cm]}

We shall consider a case when both partners
are located higher the threshold, $\pi^{\prime}~-~a_1$ system
for shortness. We can use the above formulae
with changing $\pi$ to $\pi^{\prime}$ in propagators
( but not in loops ). First of all let us discuss the
question: could not the longitudinal part obtain a scalar pole
in complex plane? In other words, would not the anzats~(\ref{anzats})
contradictory in this case ?

Let the function $D(p^2)$ (\ref{D}) now has pole in a complex plane
at the second Riemann sheet $D(\mu_c^2) = 0$. The absence  of this pole
in the longitudinal part of  vector propagator means that
 (\ref{Answer}):
\begin{equation}
    p^2 - \mu^{\prime 2} + J_{11}(p^2) = 0 \ \ \ \ \ \mbox{at} \ \ \ \
 p^2 = \mu_c^2 .
\label{1}
\end{equation}
Then from the explicit form of $D(p^2)$ there follows that the
off--diagonal loop is equal to zero as well.
\begin{equation}
    J_{12}(p^2) = 0 \ \ \ \ \ \mbox{at} \ \ \ \
 p^2 = \mu_c^2
\label{2}
\end{equation}
Let us write the more general expressions for loops as compared
with~(\ref{loops}), introducing  arbitrary subtraction
polynomials $P_{ij}$.
\begin{eqnarray}
J_{11}(p^2) = g_1^2 ( P_{11} + p^2 H(p^2) \nonumber\\
J_{12}(p^2) = g_1 g_2 ( P_{12} + H(p^2)
\end{eqnarray}
We can obtain the condition of compatibility of~(\ref{1})
and~(\ref{2}), eliminating the function $H(p^2)$ from them.
\begin{equation}
    p^2 - \mu^{\prime 2} + g_1^2 ( P_{11} - p^2 P_{12} ) = 0
\ \ \ \ \ \mbox{at} \ \ \ \
 p^2 = \mu_c^2
\end{equation}
This expression should be a second order over $p^2$ polynomial
to have a zero in  complex plane. So in this case
the anzats~(\ref{anzats}) seems to be non-- contradictory too, but
higher order of polynomials are needed as compared
with the $\pi - a_1$ case.\\[0.3cm]

\noindent \underline{The amplitude
$\pi\sigma \to ( \pi^{\prime}, a_1) \to \pi\sigma$.}\\

The calculation of this amplitude allows to look at manifestation
of S--V mixing in matrix element.
First of all it is instructive to write this amplitude with
bare propagators of $\pi^{\prime}$ and $a_1$ in s--channel.
\[
M^B = \frac{g^2_{A\pi\sigma}}{M^2 -s -i\ \epsilon}\cdot
\frac{s(t-u)+(m^2-\mu^2)^2}{s}
- \frac{g^2_{A\pi\sigma}}{M^2}\cdot \frac{(m^2-\mu^2)^2}{s}
+ \frac{g^2_{\pi^{\prime}\pi\sigma}}{\mu^2 -s -i\ \epsilon}
\]

Here the first term results from the transverse part of
$a_1$ propagator, the second one from longitudinal part
and third -- from $\pi^{\prime}$ propagator. We can see from
the angular dependence that first term has the angular momentum
J~=~1 in s--channel and the remaining ones have J~=~0.
So the longitudinal part of vector propagator gives rise to
the background nonresonance contribution to the J~=~0
amplitude. At the dressing of propagators there takes place
a unitarization of the p-- and s--wave amplitudes
separately.
As for the s--wave, there takes place a joint unitarization
in the system "pole + background".   One can suspect that
with the full propagators (~and with accounting
the off--diagonal terms~) we shall obtain the unitary expressions
both for p-- and s--wave amplitudes.
It is easy to verify with
using of our solutions~(\ref{Answer}) that this is the case
(~that's an additional check of correctness of calculations~),
but we do not present here these formulae.
So in the amplitude $\pi\sigma \to \pi\sigma$ with  J~=~0
there takes place a unitarization of resonance with the presence
of background, and the background is generated
by the longitudinal part of vector propagator.\\[0.3cm]

{\large 5.\ \ S -- V mixing and gauge bosons.\\[0.3cm]}

Let us clarify a question whether
this effect of the S--V mixing
exists for gauge bosons.  To this end we should look at the
transition loop, connecting W~(~Z~) with its scalar partner -- Higgs
particle.
\begin{equation}
J_{12}^{\alpha}(p) \sim
\int \frac{d^4 l}{(2\pi)^4}\
Sp\ \left\{ 1\cdot \frac{i}{\hat l - \hat p - m_1}\
\gamma^{\alpha} (1 - a \gamma^5)\ \frac{i}{\hat l - m_2} \right\}
\label{W}
\end{equation}
One can see that the $\gamma^5$  term does not contribute
and result is proportional to $m_1~-~m_2$, i.e. integral is
equal to zero for fermion--antifermion loop.
But if  there is the pseudoscalar vertex instead of scalar
one in (\ref{W}), then such loop is nonzero with any fermion masses.
It means that $Z^0$ boson can be connected by loop with a
pseudoscalar particle,
and  $W^{\pm}$ can transit into the charged scalar (pseudoscalar)
Higgs particle. So in the Standard Model (one neutral scalar
Higgs particle) there is no effect of unitary S--V mixing
but in extended (super--extended) variants of electroweak
models this effect can occur.\\[0.3cm]

{\large 6.\ Discussion.\\[0.5cm]}

So the existence of  loop S--V transitions
leads to a system of Dyson equations for
full propagators and the obtained full propagators
have more complicated form. Another consequence -- the appearance
of unusual transition propagators, connecting the scalar and
vector vertices.
This phenomenon, unexpected at first sight, is in fact
a very logical and has the transparent meaning. The reason is the simple
fact that vector field has an additional J~=~0 degree of freedom.

We considered in detail the renormalization of coupled propagators
for the $\pi - a_1$ system, taking into account the $\pi\sigma$
intermediate state.
That's not so real physical example but for coupled propagators
the renormalization procedure is not so evident.
In our opinion there exists the natural
physical requirement for renormalization, namely, that the longitudinal
part of vector propagator should not get the poles.
In other words the corresponding degree of freedom remains
unphysical after renormalization.
This requirement defines completely  the renormalization
procedure.

Up to now we have not touched on the question whether the considered
effect of unitary S -- V mixing leads to some physical consequences
besides the redefinition of parameters in J~=~0 system.
This question should be considered in more detail, here we shall
only say few words from general point of view.

If speaking about
effective scalar resonance (e.g. in $\pi^{\prime}~-~a_1$ system )
it's rather difficult to imagine such situation, where the
S~--~V mixing will lead to observed effects. Indeed,
from phenomenological point of view it gives a background
spin 0 contribution which will be unitarized together with
a pole. But in real life there exist different sources
of background and to observe this effect we need some special
situation: it should be rather broad resonance and background
contributions from other sources should be small.

Some peculiar situation takes place in the case of  $\pi$ -- meson
which is located under threshold of any reaction. At first sight
it seems that S -- V mixing gives only small correction
 of $m_{\pi}^2/m_{A}^2$ type to standard form of full $\pi$ -- meson
propagator because of very different masses of mixed
particles. But, firstly, there appeared the off--diagonal transitions
which were absent in a standard picture . And secondly, let us remember
that among the all $\pi$ -- meson interaction vertices there is
one vertex with the special properties -- that's the electrodynamical
one. So I suppose that it would be interesting to look at
such $\pi$ -- meson processes from the point of view
of the S~--~V mixing manifestation. The set of such processes
is well--known, see e.g. discussion of the off--shell $\pi$--meson
effects in~\cite{RFS}

It's possible that the effect under consideration takes place
for gauge bosons W, Z but only in the case of nonstandard Higgs
particles (charged or pseudoscalar). In particular the S -- V
mixing can change the pattern of the CP--violating asymmetry
between $t \to b \tau^+\nu_{\tau}$ and
$\bar t \to \bar b \tau^-\bar \nu_{\tau}$,
considered in~\cite{ass,Atwood} (~this asymmetry is proportional
to imaginary part of longitudinal propagator~). But recall that
in framework of the Standard Model there is no such effect.

In principal the S -- V mixing may be considered in arbitrary
$\xi$--gauge, where  additional scalar ghost exists,
but it will lead to more complicated picture.
Nevertheless, the above formalism
allows to look at the single vector meson dressing
in general $\xi$--gauge from another side, at least it allows
to clarify some used approximations.

In particular, at discussion in~\cite{Castro2}
it was used the general expression  from~\cite{BC} for the full
nonrenormalized vector propagator of the W in general $\xi$--gauge.
It turns out that this very complicated from first sight
expression may be written just in form~(\ref{Answer}).
That's natural because in the system "W -- ghost" there
arise the loop  scalar--vector transitions.
Then some simplifications have been made in~\cite{Castro2}, which in
fact are the following: in function D~(\ref{D}) the
term $(J_{12})^2$ is omitted since it is of the order of $g^4$ .
Evidently, after that the formulae~(\ref{Answer})
become rather trivial.
But when viewed closely, such approximation looks as unreasonable.
Indeed, in the function D there are other terms of the same  $g^4$ order
which are not omitted. Moreover we have seen  that all terms of
the  $g^4$ order are canceled with each other in D. So the approximated
formula for full vector propagator in arbitrary $\xi$--gauge
from Ref.~\cite{Castro2} looks as unjustified. But it turns out
that in the limit $\xi \to \infty$ (~unitary gauge~) the exact answer
is restored.

The similar phenomenon of unitary mixing of different spin
particles (~or fields~) should be expected at dressing of
spin 2 particle. It is described by the symmetry tensor
$T^{\mu\nu}$ with 10 degrees of freedom and only 5 of them
correspond to spin 2 particle. So at some conditions  there can
appear a loop  tensor--scalar or tensor--vector transitions.

I am indebted to V.V.Serebryakov and G.N.Shestakov for useful
discussions and to V.M.Leviant for reading the manuscript.
Special gratitude to S.Scherer for correspondence
and clarifying comments.\\[0.2cm]

{\large \ Appendix.\\[0.2cm]}

We present here the final formulae for renormalized propagators
in $\pi - a_1$ system.\\
\underline{The transverse part of vector propagator:}
\begin{eqnarray}
\Pi_{22}^T(p^2) &=& \frac{1}{M^2 - p^2 + J_{22}^T(p^2)}\nonumber\\
J_{22}^T(p^2) &=& \frac{g_{a_1\pi\sigma}^2}{48\pi}
\left[ G(p^2) - Re\ G(M^2) - (p^2 -M^2)\ Re\ G^{\prime}(M^2) \right]
\nonumber\\
G(p^2) &=& \frac{p^4}{\pi}\ \int \frac{ds}{s (s - p^2)}
\left( \frac{\lambda(s,m^2,\mu^2)}{s^2} \right)^{3/2}
\end{eqnarray}
Thus the subtraction constants in~(\ref{loop22}) are:
\begin{equation}
A(0) = \frac{g_{a_1\pi\sigma}^2}{48\pi}
\left[ Re\ G(M^2) - M^2\ Re\ G^{\prime}(M^2) \right],\
\ \ \ \ \ A^{\prime}(0) = \frac{g_{a_1\pi\sigma}^2}{48\pi} \
Re\ G^{\prime}(M^2)
\end{equation}
\underline{The coupled propagators.}\\
Loops:
\begin{eqnarray}
J_{11}(p^2) &=& g_1^2\ \left[ H(p^2) - H(\mu^2)
- (p^2 - \mu^2)\ H^{\prime}(\mu^2) \right]
\equiv g_1^2 \ (p^2 - \mu^2)^2\ H_2(p^2)\nonumber\\
J_{12}(p^2) &=& g_1\ g_2\
\left[ H(p^2) - H(\mu^2) \right]
\equiv g_1\ g_2\ (p^2 - \mu^2) H_1(p^2)\nonumber\\
J_{22}^L(p^2) &=& \tilde M^2 + g_2^2\ H(p^2),\ \ \ \ \
\tilde M^2 = M^2 + A(0) - g_2^2 H(0)
\end{eqnarray}
where
\begin{equation} H(p^2) = \frac{1}{\pi}\
\int \frac{ds}{s (s - p^2)} \left( \frac{\lambda(s,m^2,\mu^2)}{s^2}
\right)^{1/2}
\end{equation}

Full propagators are:
\begin{eqnarray}
\Pi_{11}(p^2) &=& \frac{1}{p^2 - \mu^2}\cdot
\frac{\tilde M^2 + g_2^2\ H(p^2)}{\tilde D(p^2)}
\nonumber\\
\Pi_{12}(p^2) &=&
\frac{-i\ g_1 g_2\ H_1(p^2)}{\tilde D(p^2)}
\nonumber\\
\Pi_{22}^L(p^2) &=&
\frac{1 + g_1^2\ (p^2 - \mu^2)\ H_2(p^2)}{\tilde D(p^2)},
\end{eqnarray}
where
\[
D(p^2) = (p^2 - \mu^2)\ \tilde D(p^2)
\]
Let us recall the notations:
\[
g_1^2 = \frac{g_{\sigma\pi\pi}^2}{16 \pi} \ \ \ \ \ \ \ \
g_2^2 = (m^2 - \mu^2)^2 \cdot \frac{g_{a_1\pi\sigma}^2}{16 \pi}
\]

\end{document}